# Carrier induced ferromagnetism in room temperature ferromagnetic semiconductor rutile $TiO_2$ doped with Co

H. Toyosaki,[1] T. Fukumura,[1,†,§] Y. Yamada,[1] K. Nakajima,[2,3] T. Chikyow,[2,3] T. Hasegawa,[3,4] H. Koinuma,[3,5] M. Kawasaki[1,3]

[1]*Institute for Materials Research, Tohoku University, Sendai 980-8577, Japan.*
[2]*National Institute for Materials Science, Tsukuba 305-0044, Japan.*
[3]*Combinatorial Materials Exploration and Technology, Tsukuba 305-0044, Japan.*
[4]*Department of Chemistry, University of Tokyo, Tokyo 113-0033, Japan.*
[5]*Materials and Structures Laboratory, Tokyo Institute of Technology, Yokohama 226-8503, Japan.*

A ferromagnetic semiconductor, rutile $(Ti,Co)O_2$, exhibits anomalous Hall effect at room temperature. Strong dependence of the anomalous Hall effect, as well as the magneto-optic response, on the carrier concentration suggests carrier induced ferromagnetism in this compound. Both ferromagnetic responses are caused by the charge carriers at the band edge of host semiconductor, indicating possibilities of spintronics devices operable at room temperature.

Ferromagnetic semiconductors have attracted much attention because they can serve for spintronics devices utilizing charge and spin degrees of freedom simultaneously.[1,2] The most extensively studied ferromagnetic semiconductor, (Ga,Mn)As, was used to demonstrate that the Curie temperature $T_C$ can be controlled by electric field, promising for the integration of spintronics device into electronic circuits.[3] Charge carriers in ferromagnetic semiconductors should be spin polarized, as has been demonstrated by circularly polarized light emission or spin dependent tunneling.[4,5] Therefore, numbers of application can be explored by utilizing these features of ferromagnetic semiconductors. However, $T_C$ of this system is limited to 160 K.[6] The most important breakthrough in this field is to elevate $T_C$ well beyond room temperature in order to realize practical devices operable at room temperature. Therefore, various host semiconductors and dopant transition metal elements have been tested for realizing room temperature ferromagnetic semiconductors.

Diluted magnetic oxide, oxide semiconductor doped with transition metal ions, is one of the most promising candidates for ferromagnetic semiconductor having higher $T_C$.[7,8] Recently, there have been large amount of reports on diluted magnetic oxides claiming higher $T_C$ than (Ga,Mn)As, although the ferromagnetic origin for these compounds is still a subject of intense debate.[9] A transparent oxide semiconductor (Ti,Co)$O_2$ was found to be ferromagnetic well above room temperature for both anatase and rutile phases.[10,11] Several studies, however, have criticized that the ferromagnetism is caused by the precipitations of ferromagnetic Co clusters.[12,13] In order to prove that the ferromagnetism is induced not by ferromagnetic precipitation but by intrinsic mechanism, ferromagnetic responses that are caused by charge carriers are more convincing evidence than magnetization measurement.

Here, we report on the observation of ferromagnetic anomalous Hall effect and ferromagnetic magneto-optical effect for rutile (Ti,Co)$O_2$ with different carrier concentrations at room temperature. The both effects represent ferromagnetic responses from charge carriers at the band edges of host semiconductor. This compound shows strong dependence of the ferromagnetic responses on the charge carrier concentration, suggesting that the ferromagnetic origin is carrier induced mechanism.

Rutile $Ti_{0.97}Co_{0.03}O_2$ epitaxial thin films were fabricated by using laser molecular beam epitaxy. $Ti_{0.97}Co_{0.03}O_2$ ceramics target was ablated by KrF excimer laser.

The films with (101) orientation were grown on *r*-sapphire substrates at 700°C. Oxygen pressure during the growth was varied as $1 \times 10^{-8}$ Torr (sample #1), $1 \times 10^{-6}$ Torr (sample #2), and $1 \times 10^{-4}$ Torr (sample #3) in order to control *n*-type carrier concentration by introducing controlled concentration of oxygen deficiencies. Typical film thickness was 60-80 nm. From x-ray diffraction, scanning electron microscopy, transmission electron microscopy, and atomic force microscopy, neither impurity phase nor precipitation was observed. Absorption spectra were deduced from transmission and reflection measurements. Magnetic circular dichroism (MCD: difference of absorption in a medium between left and right circularly polarized lights) spectra were measured with transmission configuration, where the details of the measurement were described elsewhere.[14] Photolithographically patterned Hall bars (200 µm long × 60 µm wide) were used for electronic transport measurements, where magnetic field was applied along out-of-plane.

As seen in the inset table of Fig. 1(a), the resistivity $r_{xx}$ at 300 K can be controlled by orders of magnitude due to effective *n*-type carrier doping through low oxygen pressure growth. Figure 1(a) shows the absorption spectra for the samples #1-#3. The absorption edges for the samples #2 and #3 are almost identical at about 3.0 eV, whereas the absorption edge for the sample #1 is shifted to 3.5 eV. Figure 1(b) shows MCD spectra in a magnetic field of 1 T for these samples. The insulative sample #3 shows negligible MCD signal representing the negligible magnetization, while the samples #1 and #2 show large MCD signal from visible to ultraviolet region having ferromagnetic magnetization (see Fig. 2(b)). The MCD signal is negative below ~3 eV and positive above ~3 eV with a peak structure, whose behavior is similar to that of the anatase $(Ti,Co)O_2$.[14] It is noted that the energy shift in positive MCD peak in Fig. 1(b) is almost equal to the blue shift of absorption edge in Fig. 1(a) from the sample #2 to #1. The MCD signal at the absorption edge of magnetic semiconductors is originated from the Zeeman splitting of charge carriers at the band edge.[15] Therefore, huge magneto-optical response tracing the shift of absorption edge is a manifestation of magnetic response due to charge carriers interacting with magnetic impurities. The blue shift of absorption edge could be originated from the increase of carrier concentration via like Burnstein-Moss effect, although the blue shift of ~0.5 eV may be too large to be explained by simple parabolic band model.[16]

Figure 2(a) shows magnetization hysteresis for the sample #2, where magnetic field is applied along out-of-plane (solid line) and in-plane (dashed line), respectively. These results indicate that the easy magnetization axis is along out-of-plane.[17] One of the previous papers pointed out that isotropic magnetization for anatase (Ti,Co)O$_2$ is an indication of Co precipitation.[13] The observed anisotropy in magnetization can be an experimental evidence to rule out the ferromagnetic precipitations with isotropic magnetization. Figure 2(b) shows MCD hysteresis for the samples #1 and #2 measured at photon energies of 3.9 eV and 3.4 eV, respectively. Both signal shows ferromagnetic behavior and such ferromagnetic hysteresis is also observed for the different photon energy, at which MCD signal is finite (not shown). The magnetic field dependence of magnetization and MCD for the sample #2 shows very similar behavior, representing that both measurements probe the same intrinsic magnetization arisen from the net magnetization of the film and the magneto-optical response of carriers at absorption edge, respectively.

Here, we demonstrate ferromagnetic response of charge carriers through Hall effect, where the Hall effect is emergence of voltage transverse to the applied current and external magnetic field. Generally, Hall effect of ferromagnets is expressed as $r_{xy} = R_o B + R_s \mu_0 M$, where $B$ is the magnetic induction, $\mu_0$ is the magnetic permeability, $M$ is the magnetization, $R_o$ is the ordinary Hall coefficient, and the $R_s$ is the anomalous Hall coefficient. The first term denotes ordinary Hall effect and the second term denotes the anomalous Hall effect, where the latter term dominates over the former term in typical ferromagnets. If the charge carrier is ferromagnetically spin polarized, the Hall resistivity $r_{xy}$ vs. $B$ ($B \sim \mu_0 H$) curve should behave like the magnetization hysteresis curve. Figure 3(a) shows magnetic field dependence of $r_{xy}$ at different temperatures for the sample #1. Indeed, $r_{xy}$ at 300 K increases abruptly with increasing magnetic field from 0 to 0.5 T, and almost saturates over 1 T, that is coincident with the magnetization and MCD behaviors shown in Fig. 2. Thus, this abrupt increase of $r_{xy}$ is evidently attributed to the anomalous Hall effect caused by ferromagnetic spin polarization of charge carrier, the behavior of which being different from that of Co metal.[18] Contrasted with Fig. 3(a), overall feature of $r_{xy}$ for the less conductive sample #2 in Fig. 3(b) is proportional to magnetic field representing appearance of ordinary Hall effect at different temperature. The sign of the slope indicates $n$-type carrier and the change in slope with temperature indicates the decrease of carrier concentration with decreasing temperature as shown in the inset table. By extracting ordinary part $r_o$ ($\propto H$) from $r_{xy}$, however, the nonlinear terms appear as can be seen in the inset of Fig. 3. The sample #2 also shows anomalous Hall effect similar to the sample #1, although the ordinary Hall effect is dominant for the sample #2.

   We shall discuss further the anomalous Hall effect shown in Fig. 3. There are two features in $r_{xy}$ (- $r_o$) commonly observed: (i) the increase with increasing magnetic field and (ii) the increase with decreasing temperature shown in Fig. 3(a) and inset of Fig. 3. For the former (i), the positive slope in $r_{xy}$ could be interpreted as the $p$-type majority carrier from ordinary Hall effect. However, the sample #1 has to contain more $n$-type carrier than the sample #2 because of the low resistivity due to more oxygen deficiency. Therefore, the positive slope does not mean the $p$-type majority carrier although the origin is unknown at present. On the other hand, the latter (ii) cannot be explained by the increase of magnetization because the magnetization of this compound is almost constant below 300 K. Figure 4(a) shows temperature dependence of $r_{xx}$ and $r_{xy}$ for both samples. Both $r_{xx}$ and $r_{xy}$ increase with decreasing temperature, and $r_{xy}$ is approximately proportional to $r_{xx}$ below 300 K. Accordingly, the increase of $r_{xy}$ with decreasing temperature can be interpreted as an appearance of skew scattering (anisotropic scattering of charge carrier by impurities in the presence of spin-orbit coupling) at low temperature, that has been observed in diluted magnetic alloys[19] possibly representing the impurity scattering of charge carrier by Co ion.

   In general, skew scattering is a phenomenon observed at low temperature. Therefore, the increase of $r_{xy}$ with decreasing temperature can be explained by skew scattering, but the anomalous effect at 300 K could not be explained because of the large thermal disturbance. Recently, the anomalous Hall effect of ferromagnets, that has been unsolved problem for long decades, is explained in the context of Berry phase acquired by carrier: whose effect on the carrier motion is equivalent to the phase caused by a magnetic field.[20-22] Ye et al.[20] and Taguchi et al.[21] attribute the anomalous Hall effects of colossal magneto-resistive ferromagnet (La,Ca)MnO$_3$ and spin frustrated ferromagnet Nd$_2$Mo$_2$O$_7$, respectively, to Berry phase obtained by noncolinear spin configuration, and Jungwirth *et al.*[22] attributes that of III-V ferromagnetic semiconductors such as (Ga,Mn)As to Berry phase caused by spin-split Fermi surface. The latter calculates the dependence of anomalous Hall conductivity on carrier concentration quantitatively. Let us reexamine the data of Fig. 3 in order to compare with the result of Ref. 22. Since the temperature dependence of mobility is much smaller than that of carrier concentration as seen in the inset table of Fig. 3(b), temperature dependence of $r_{xx}$ can be approximately regarded the change of carrier concentration. Figure 4(b) shows the relationship between Hall conductivity $s_{xy}$ and conductivity $s_{xx}$ replotted from Fig. 4(a). The monotonous increase of $s_{xy}$ with $s_{xx}$ for wide range of $s_{xx}$, that is proportional to carrier concentration, in different samples implies a universal behavior of $s_{xy}$ on carrier concentration in this compound. In comparison with (Ga,Mn)As, $s_{xy}$ is much smaller. This is possibly because of the small spin orbit coupling and/or small exchange interaction between electron carrier and spin in this compound.

In conclusion, the magneto-optical response and anomalous Hall effect strongly dependent on carrier concentration suggest carrier induced ferromagnetism in this compound. Thus, it should be possible to control the ferromagnetic properties of semiconductor via carrier modulation at room temperature, for example, through gate bias voltage in a configuration of field effect transistor or photo excitation. In addition, the anomalous Hall effect seen in this compound would provide a case study different from a ferromagnetic semiconductor (Ga,Mn)As.

**Acknowledgements**

We acknowledge K. Ando for magneto-optical measurement and discussion. This work was supported by the Ministry of Education, Culture, Sports, Science and Technology, Grant-in-Aid for Creative Scientific Research (14GS0204), the Inamori Foundation, and the Murata Science Foundation.

**Figure captions**

FIG. 1. (a) The absorption spectra of ultraviolet-visible region measured at 300 K for $Ti_{0.97}Co_{0.03}O_2$ epitaxial thin films (samples #1, #2, and #3) grown under various oxygen pressures, where the samples #1 and #2 are indicated as solid lines and the sample #3 is indicated as dashed line. The inset table shows the relationship between the oxygen pressure during growth ($P_{O2}$) and resistivity ($r_{xx}$) at 300 K for these samples. (b) The magnetic circular dichroism (MCD) spectra measured in a magnetic field of 1 T at 300 K.

FIG. 2. (a) The magnetization hysteresis measured at 300 K for the sample #2. Solid and dashed lines denote out-of-plane and in-plane magnetizations, respectively. (b) The MCD hysteresis with out-of-plane magnetization at 300 K at photon energies of 3.4 eV and 3.9 eV for the samples #1 and #2, respectively.

FIG. 3. The magnetic field dependence of Hall resistivity $r_{xy}(H)$ at 130 K (square symbol), 200 K (triangle symbol), and 300 K (circle symbol) for the sample #1 in (a) and #2 in (b). The inset shows the anomalous part of Hall resistivity $r_{xy}(H) - r_o(H)$ for the sample #2 deduced by extracting the ordinary part of Hall resistivity $r_o(H)$ that is linear against magnetic field. From $r_o(H)$, carrier concentration ($n$) and mobility ($m$) are evaluated and listed in the inset table.

FIG. 4. (a) The temperature dependences of $r_{xx}(0\ T)$ and $r_{xy}(1\ T)$ for the samples #1 and #2, where the samples #1 and #2 for $r_{xy}(1\ T)$ are square and circle symbols, respectively. $r_{xy}(1\ T)$ for the sample #2 corresponds to the anomalous part of Hall effect, $r_{xy}(1\ T) - r_o(1\ T)$, as defined in the inset of Fig. 3. (b) The relationship between $s_{xx}$ and $s_{xy}$ for the samples #1 (square symbol) and #2 (circle symbol) replotted from (a), where $s_{xx} \sim r_{xx}(0\ T)^{-1}$ and $s_{xy} = r_{xy}(1\ T)/(r_{xx}(0\ T)^2 + r_{xy}(1\ T)^2)$.

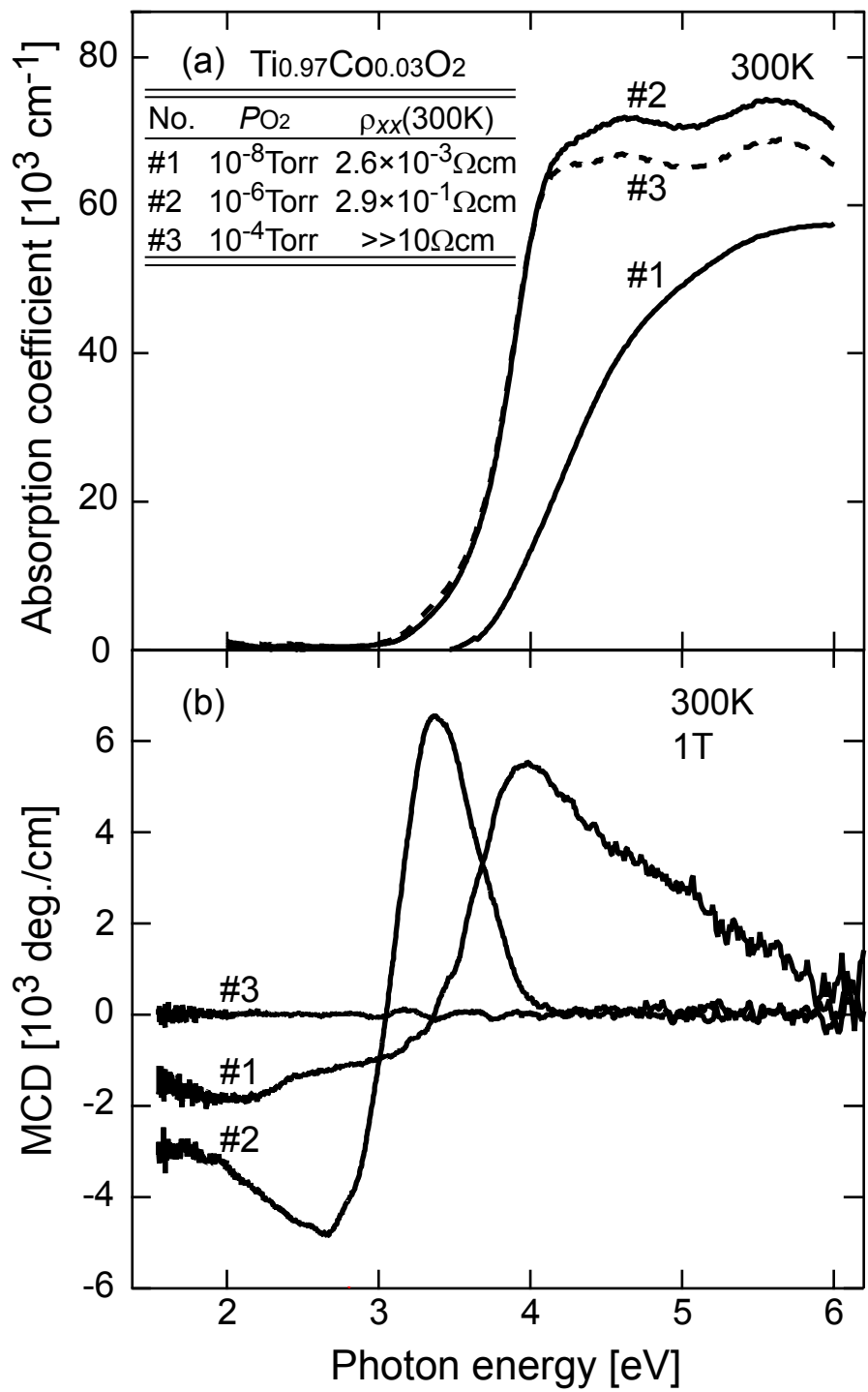

Fig. 1 H. Toyosaki et al.

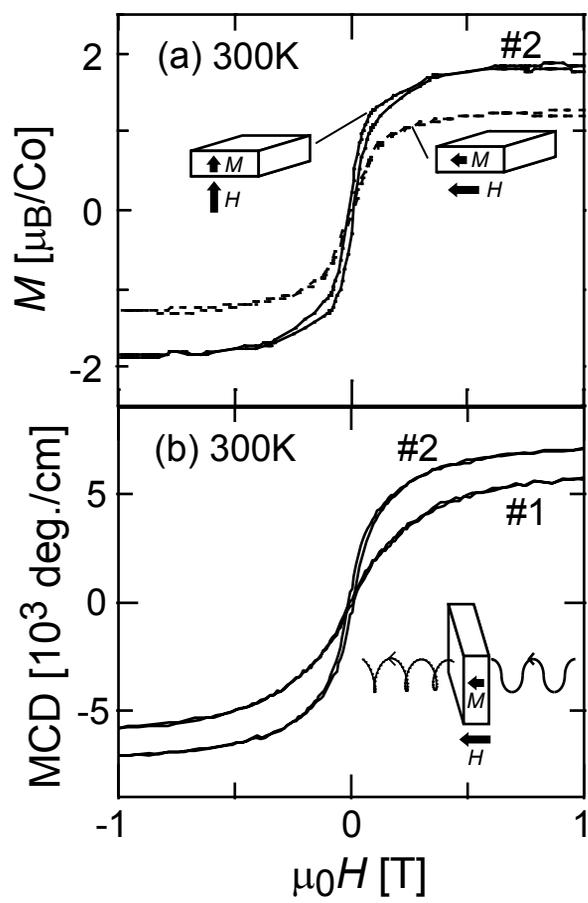

Fig. 2  H. Toyosaki et al.

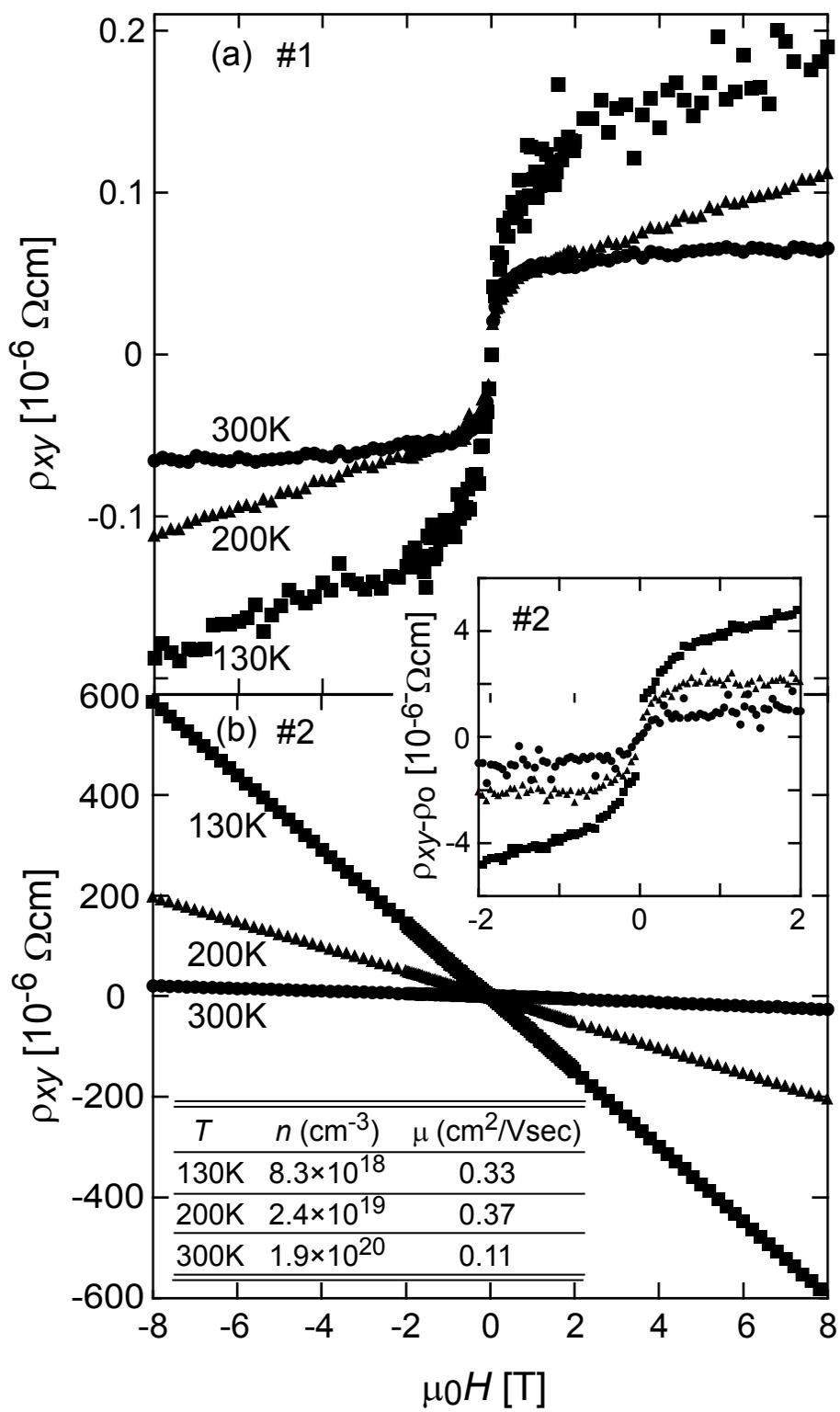

Fig. 3 H. Toyosaki et al.

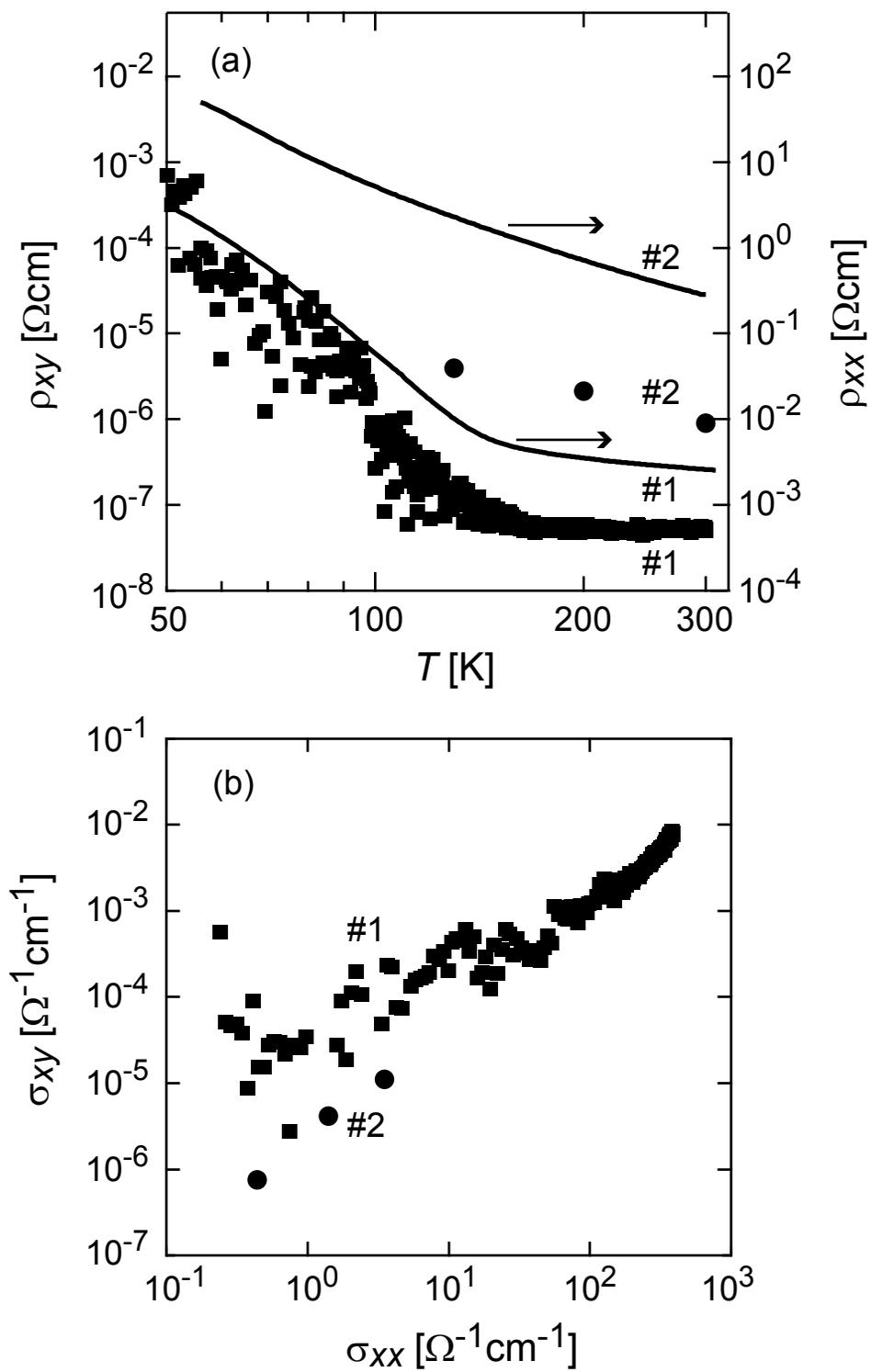

Fig. 4  H. Toyosaki et al.